\documentclass[final,5p,times,twocolumn]{elsarticle}

\usepackage{graphicx}
\usepackage{array}
\usepackage[dvips,colorlinks,linkcolor=red,anchorcolor=blue,citecolor=green]{hyperref}

\makeatletter
\long\def\@makecaption#1#2{%
  \vskip\abovecaptionskip
  \sbox\@tempboxa{#1\  #2}%
  \ifdim \wd\@tempboxa >\hsize
    #1\ #2\par
  \else
    \global \@minipagefalse
    \hb@xt@\hsize{\hfil\box\@tempboxa\hfil}%
  \fi
  \vskip\belowcaptionskip}
\makeatother
\usepackage{amssymb}
\usepackage{amsmath}

\journal{arXiv.org}

\begin{document}

\begin{frontmatter}

\title{Investigation of target in C-ADS and IAEA ADS benchmark \tnoteref{lab1}}
\tnotetext[lab1]{Supported by National Natural Science Foundation of China (11045003 \& 10975150)}
\author[label1,label2]{Guojun Hu \fnref{lab2}}
\fntext[lab2]{huguojun@mail.ustc.edu.cn}
\author[label1,label2]{Hongli Wu}
\author[label1,label4]{Tian Jing}
\author[label1,label2]{\\Xiangqi Wang\fnref{lab3}}
\fntext[lab3]{Corresponding author. wangxaqi@ustc.edu.cn}
\author[label3,label1]{Jingyu Tang}
\address[label1]{National Synchrotron Radiation Laboratory,USTC}
\address[label2]{School of Nuclear Science and Technology, USTC}
\address[label3]{Institute of High Energy Physics Chinese Academy of Sciences}
\address[label4]{Department of Modern Physics, USTC}

\begin{abstract}
The spatial and energy distribution of spallation neutrons have an effect on the performance of Accelerator Driven Subcritical systems. In this work, the spatial, energy distribution of spallation neutrons and the effect of these factors on proton efficiency was studied. When the radius of spallation region increases, backward neutrons were found to have a rather big ratio and have a positive effect on proton efficiency. By making better use of these neutrons, we may increase the radius of target to satisfy some other requirements in the design of subcritical core.
\end{abstract}


\end{frontmatter}

\section{Introduction}
\label{sec1}
China ADS (C-ADS) consists of a high-intensity proton accelerator in CW (continuous wave) mode with proton energy of 0.6 GeV in phase three and 1.5 GeV in phase four, a subcritical core, and a spallation target of liquid Lead-Bismuth Eutectic (LBE or Pb-Bi), which is favored in projects for its low melting point, high boiling point and low neutron absorption cross section \cite{Bauer}.

The target should have such a size that it incepts the main part of the high-energy cascade and then we get a optimized neutron multiplicity\cite{Krasa}. Neutron multiplicity is defined to be the number of neutrons produced per beam particle. Increasing the size of target will affect the spatial and energy distribution of spallation neutrons and thus affecting the performance of the total system. So in this work, we discussed the effect of the spatial and energy distribution of spallation neutrons.

Neutron multiplicity, spatial and energy distribution were simulated with FLUKA \cite{Ferrari,Battistoni} in this work. In these simulations, Intra-nuclear cascade, pre-equilibrium, evaporation and fission models were activated. Proton efficiency of the system was partly simulated with RMC(Reactor Monte Carlo code)\cite{RMC1,RMC2}. IAEA ADS benchmark \cite{Slessarev} was an ADS model and was used in these simulations, shown in \hyperref[Fig:2]{Fig. \ref*{Fig:2}}.
\section{Validation of FLUKA}
\label{sec2}
Double-differential distribution of spallation neutrons were measured in the Laboratoire National Sturne experiment \cite{Ledoux}. In order to observe the availability of FLUKA in simulating spallation reaction, this experiment was simulated with FLUKA.
\begin{table*}[tp]
\centering
  \setlength{\abovecaptionskip}{0pt}
  \setlength{\belowcaptionskip}{0pt}
\caption{Parameters of Saturne experiment\cite{Ledoux}}\label{Tab:1}
\begin{tabular}{ll|ll|ll}
  \hline
  \multicolumn{2}{c|}{Beam} & \multicolumn{2}{c|}{Target}  &\multicolumn{2}{c} {Detectors}\\
  \hline
  Projectile   & Proton  & Material     & Lead              & $R$       & 800 cm \\
  $E_{\rm{p}}$ & 1.6 GeV & $R_{\rm{s}}$ & 1.5 cm            & $dS$      & 1963.5 $\rm{cm}^{2}$ \\
  Direction    & (0,0,1) & $N$          & $3.30\times10^{28} \rm{m^{-3}}$ & $d\Omega$ & $3.068\times10^{-3} \rm{sr}$ \\
  $R_{\rm{b}}$ & 1.4 cm  & $L_{\rm{s}}$ & 2.0 cm            &  & \\
  \hline
\end{tabular}
\end{table*}
\begin{figure}[hbt]
  \centering
  \setlength{\abovecaptionskip}{-10pt}
  \setlength{\belowcaptionskip}{-10pt}
  \setlength{\intextsep}{-20pt plus 0pt minus 0pt}
  \includegraphics[width=0.5\textwidth]{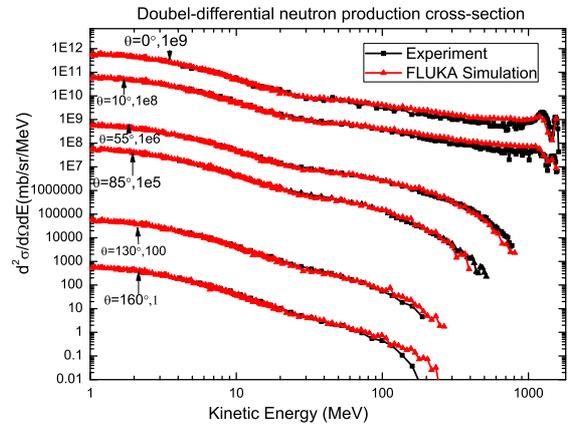}\\
  \caption{Double-differential neutron production cross-section. Data of different polar angle had  been scaled with different ratio marked in the figure. $1\sigma<5\%$ } \label{Fig:1}
\end{figure}
Parameters of target, beam and detectors in this simulation are listed in \hyperref[Tab:1]{Table \ref*{Tab:1}}. Detectors are in a sphere surface with radius of $R=800\ \rm{cm}$. Detectors are set in polar angle of $\theta=0^{\circ},10^{\circ},55^{\circ}$,$85^{\circ},130^{\circ},160^{\circ}$.
Double-differential cross-section of neutron production is obtained with Eq.(\ref{eq1}):
\begin{equation}\label{eq1}
 \frac{{\rm d}^{2}\sigma}{{\rm d}\Omega {\rm d}E} =\frac{{\rm d}N_{\rm{ns}}}{N_{\rm{primary}}\frac{{\rm d}S}{R^{2}}{\rm d}E}\frac{1}{NL_{\rm{s}}}=\frac{{\rm d}N_{\rm{ns}}}{N_{\rm{primary}}{\rm d}S{\rm d}E}R^{2}\frac{1}{NL_{\rm{s}}}
\end{equation}
Where, ${\rm d}S$ , ${\rm d} \Omega={\rm d}S/R^2$ , $N$ and $L_{\rm{s}}$ is area of detector, solid angle, atomic density and thickness of target in \hyperref[Tab:1]{Table \ref*{Tab:1}}. ${\rm d}N_{\rm{ns}}/({N_{\rm{primary}}{\rm d}S{\rm d}E})$ is scored with USRBDX card.

Comparison between this simulation and the experiment is shown in \hyperref[Fig:1]{Fig. \ref*{Fig:1}}. Data of the experiment was obtained from EXFOR/CRISRS. The result of FLUKA agrees well with the experiment. There is difference in high energy due to the defect of physical models and the statistic error of simulation. FlUKA is suitable to simulate the spallation reaction.
\section{Description of ADS benchmark and proton efficiency}
\label{sec3}
\subsection{Description of IAEA ADS benchmark}
\label{sec31}
\begin{figure}[hbt]
  \centering
  \setlength{\abovecaptionskip}{-10pt}
  \setlength{\belowcaptionskip}{-5pt}
  \includegraphics[width=0.5\textwidth]{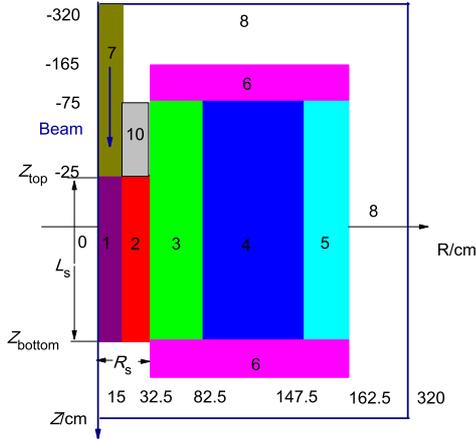}\\
  \caption{RZ view of IAEA ADS benchmark. The different regions are:(1)Target; (2)Buffer; (3)Core1; (4)Core2; (5)Core3; (6)Reflector; (7)Beam pipe; (8),(10)Moderator}\label{Fig:2}
\end{figure}

IAEA ADS benchmark was chosen to simulate the effect of spallation neutrons. Geometry and parameters of IAEA ADS benchmark is show in \hyperref[Fig:2]{Fig. \ref*{Fig:2}} and \hyperref[Tab:2]{Table \ref*{Tab:2}}. In regions "core1" and "core2", the ratio between 233-U and 232-Th is 1:9 and effective multiplication factor is: $k_{\rm{eff}}=0.96455\pm0.00062$.
\begin{table*}[tp]
  \centering
  \caption{Parameters of ADS benchmark\cite{Slessarev}}\label{Tab:2}
  \begin{tabular}{llllll}
     \hline
     \multicolumn{6}{c}{The element of IAEA ADS benchmark at BOL (20 $^\circ$C) $10^{24}\rm{cm}^{-3}$}\\
     \hline
     Nuclides      & Core1 & Core2 & Core3 & Reflector & Moderator, Target, Buffer \\
     232-Th        &       &       & 7.45 &  &  \\
     232-Th,233-U & 6.350 & 7.450 &  &  &       \\
     O  & 12.70 & 14.90 & 14.90 &  &            \\
     Fe & 8.100 & 8.870 & 8.870 & 6.630 &       \\
     Cr & 1.120 & 1.060 & 1.060 & 0.800 &       \\
     Mn & 0.046 & 0.051 & 0.051 & 0.038 &       \\
     W  & 0.046 & 0.051 & 0.051 & 0.038 &       \\
     Pb & 17.70 & 15.60 & 15.60 & 0.024 & 30.50 \\
     \hline
     \multicolumn{6}{c}{The source of IAEA ADS benchmark}\\
     \hline
     \multicolumn{3}{l}{Projectile}&\multicolumn{3}{l}{Proton}\\
     \multicolumn{3}{l}{Kinetic Energy:$E_{\rm{p}}$}&\multicolumn{3}{l}{1.0 GeV}\\
     \multicolumn{3}{l}{Beam radius:$R_{\rm{b}}$}&\multicolumn{3}{l}{10.0 cm}\\
     \hline
   \end{tabular}
\end{table*}
Liquid Lead is the material of moderator and target in ADS benchmark. Spallation region is bounded by the subcritical core and beam pipe. We divided the spallation region into "target" and "buffer" as used in "Energy Amplifier"\cite{Rubbia}. $R_{\rm{s}}$ and $L_{\rm{s}}$ is the size of spallation region, see \hyperref[Fig:2]{Fig. \ref*{Fig:2}}. In simulations of FLUKA, except for region "target", all other regions were filled with vacuum.
\subsection{Neutron source efficiency and proton efficiency}
\label{sec32}
Neutron flux  $\phi_{\rm{s}}$ in the subcritical core is the solution to the inhomogeneous steady-state neutron transport equation Eq.(\ref{eq2}):
\begin{equation}\label{eq2}
  A\phi_{\rm{s}}=F\phi_{\rm{s}}+S_{\rm{n}}
\end{equation}
Where $F$ is fission operator, $A$ is net neutron loss operator and $S_{\rm{n}}$ is external neutron source.\\
The subcritical multiplication factor $k_{\rm{s}}$ \cite{Shahbunder} is defined as Eq.(\ref{eq3}):
\begin{equation}\label{eq3}
  k_{\rm{s}}=\frac{<F\phi_{\rm{s}}>}{<F\phi_{\rm{s}}>+<S_{\rm{n}}>}
\end{equation}
$k_{\rm{s}}$ describes the number of fission neutrons produced per lost neutron in the subcritical system.\\
Neutron source efficiency, usually denoted as $\varphi^{*}$ \cite{Shahbunder}, represents the efficiency of the external source neutrons and can be expressed as Eq. (\ref{eq4}):
\begin{equation}\label{eq4}
  \varphi^{*}=(\frac{1}{k_{\rm{eff}}-1})\frac{<F\phi_{\rm{s}}>}{S_{\rm{n}}}
\end{equation}
$<F\phi_{\rm{s}}>$ is total number of neutrons produced by fission. $<S_{\rm{n}}>$ is total number of external source neutrons. For a given values of $k_{\rm{eff}}$ and $<S_{\rm{n}}>$, the larger $\varphi^{*}$, the bigger the fission power. Definition of neutron source efficiency needs the definition of source neutrons $<S_{\rm{n}}>$. In this work, $<S_{\rm{n}}>$ means neutrons escaping the spallation region.\\
Relationship between $k_{\rm{keff}}$ and $k_{\rm{s}}$ \cite{Shahbunder} is:
\begin{equation}\label{eq5}
  (1-\frac{1}{k_{\rm{keff}}})=\varphi^{*}(1-\frac{1}{k_{\rm{s}}})
\end{equation}
Proton efficiency\cite{Seltborg}, denoted as $\psi^{*}$, is defined to be:
\begin{equation}\label{eq6}
  \psi^{*}=\varphi^{*}\frac{<S_{\rm{n}}>}{<S_{\rm{p}}>}=(\frac{1}{k_{\rm{eff}}-1})\frac{<F\phi_{\rm{s}}>}{<S_{\rm{p}}>}
\end{equation}
$S_{\rm{p}}$ is total number of source protons.

To study proton efficiency of the system, simulations had been performed in two steps:
\begin{itemize}
  \item protons impact on the target, neutron multiplicity and neutrons escaping the target are recorded with FLUKA.
  \item the source neutrons recorded in the first step are then processed as an external neutron source and transported in the system.
\end{itemize}
In order to reduce the time of simulation, the height and diameter of the total system was cut to be 360 cm.
\section{Spatial and energy distribution of spallation neutrons}
\label{sec4}
\subsection{Description of spatial and energy distribution of spallation neutrons}
\label{sec41}
Spatial and energy distribution of spallation neutrons have an effect on proton efficiency. 232-Th,238-U, Am can be used as fuels of ADS. An external neutron hardly ever induce fissions of these fuels when its kinetic energy is below 1 MeV\cite{Seltborg}. The energy spectrum of neutrons in subcritical core is different from that of spallation neutrons; the former is determined by geometry and composition of the system, while the later is mostly determined by the size of spallation region. The spatial distribution of spallation neutrons determines the distribution of neutron flux in the subcritical core and thus determining the proton efficiency of the system.
\begin{figure}[hbtp]
  \centering
  \setlength{\abovecaptionskip}{-10pt}
  \setlength{\belowcaptionskip}{-10pt}
  \includegraphics[width=0.5\textwidth]{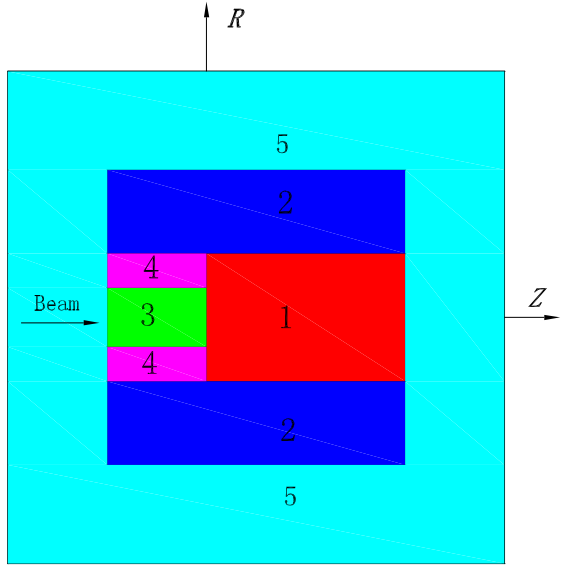}\\
  \caption{RZ view of target model used in FLUKA. The different regions are: (1)Target; (2)Subcritical core; (3)Beam pipe; (4)Moderator; (5)Void. In simulations of FLUKA, region Moderator and Subcritical core is filled with vacuum. Spallation neutrons include : Backward(1$\rightarrow$3,4); Forward(1$\rightarrow$5); Radial(1$\rightarrow$2); Survival(4$\rightarrow$2); Leak(3,4$\rightarrow$5)}\label{Fig:3}
\end{figure}
In this work, spatial and energy distribution of spallation neutrons was simulated with FLUKA. The target model used in FLUKA is shown in \hyperref[Fig:3]{Fig. \ref*{Fig:3}}. Neutrons are divided into forward, radial, backward neutrons according to the surface the neutrons cross when they escape the target. Radial neutrons enter the subcritical core directly. Forward neutrons hardly ever enter the subcritical core. Some backward neutrons make it to the subcritical core through the moderator between the beam pipe and the subcritical core (region "10" in \hyperref[Fig:2]{Fig. \ref*{Fig:2}}), called survival neutrons; other backward neutrons enter the reflector or leak out from the beam pipe, called leak neutrons. A clear explanation of these neutrons is shown in \hyperref[Fig:3]{Fig. \ref*{Fig:3}}. Except for region "target", all other regions are filled with vacuum.

As for spatial distribution, what we concern here was the neutron current surface density $I_{\rm{r}}(Z)$ in the side surface of target. $Z$ is the distance between a point in the side surface and the top surface of target. We also cared about the number of backward, forward and radial neutrons. As for energy distribution, the energy spectrum of radial neutrons were mostly discussed.

\subsection{Spatial and energy distribution of radial neutrons in C-ADS}
\label{sec42}
Radial neutrons are the main part of spallation neutrons and their spatial distribution is closely related to the neutron flux in the subcritical core. So we studied this factor quantitatively with FlUKA. Remember that we mainly care about neutron current surface density here.

In this simulation, proton source is a circular uniform beam with radius of $R_{\rm{b}}=15\ \rm{cm}$ in C-ADS HEBT (High Energy Beam Transport ) \cite{HEBT}. Simulation had been performed with parameters listed in \hyperref[Tab:3]{Table \ref*{Tab:3}}. To obtain the neutron current surface density, subroutine mgdraw.f was modified and activated to record neutrons escaping the "target" region in \hyperref[Fig:3]{Fig. \ref*{Fig:3}}. Then we constructed the spatial and energy distribution of neutrons with the information recorded. The result of this simulation is shown in \hyperref[Fig:4]{Fig. \ref*{Fig:4}}. Average kinetic energy of spallation neutrons in different position was used in calculating energy distribution: $\overline{E_{\rm{r}}(Z)}$.
\begin{table}[hbtp]
  \centering
  \setlength{\abovecaptionskip}{-5pt}
  \caption{Parameters of target and beam of C-ADS \cite{HEBT}}\label{Tab:3}
  \begin{tabular}{ll|ll}
    \hline
    \multicolumn{2}{c|}{Target characteristic} & \multicolumn{2}{c}{Beam}\\
    \hline
    Material & 50\%Pb,50\%Bi & Projectile & Proton \\
    $R_{\rm{s}}$ & 25.0 cm   & $E_{\rm{p}}$  & 1.5 GeV \\
    $L_{\rm{s}}$ & 100.0 cm  & Source & HEBT \\
    \hline
  \end{tabular}
\end{table}
As can be seen in \hyperref[Fig:4]{Fig. \ref*{Fig:4}}, as $Z$ increase: $\overline{E_{\rm{r}}(Z)}$ reaches a minimum and then saturates; $I_{\rm{r}}(Z)$ reaches a maximum and then decreases to zero. Based on this result, if the position of maximum $I_{\rm{r}}(Z)$ locates at the axial center of subcritical core(that is $z=0\ \rm{cm}$), neutron flux $\phi_{\rm{s}}(z)$ in the subcritical core will be symmetric (with axis at $z=0\ \rm{cm}$). Generally, an symmetric distribution of $\phi_{\rm{s}}(z)$ is a key factor in designing the system. In this way, we may optimize the position(relative to the subcritical core) of spallation target, discussed in \hyperref[sec441]{Sec. \ref*{sec441}}.
\begin{figure}[hbt]
  \centering
  \setlength{\abovecaptionskip}{-10pt}
  \setlength{\belowcaptionskip}{-10pt}
  \setlength{\intextsep}{-10pt plus 0pt minus 0pt}
  \includegraphics[width=0.5\textwidth]{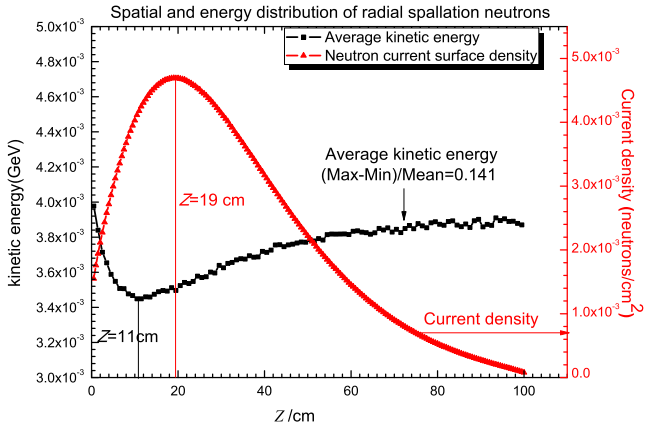}
  \caption{Neutron current surface density $I_{\rm{r}}(Z)$ and average kinetic energy$\overline{E_{\rm{r}}(Z)}$ as a function of $Z$. $I_{\rm{r}}(Z)$ is normalized to per source proton.}\label{Fig:4}
\end{figure}
In this case, $\overline{E_{\rm{r}}(Z)}$ changes slightly, $(E_{\rm(rmax)}-E_{\rm{rmin}})/E_{\rm(raverage)}\approx0.141$. Energy distribution of radial spallation neutrons has a influence on the neutron flux of the subcritical core, but it is a less important factor than the spatial distribution of radial neutrons. Detailed effect of the energy distribution needs further research.

\subsection{Backward neutrons and radial neutrons as a function of $R_{\rm{s}}$}
\label{sec43}
\begin{figure}[hbt]
  \centering
  \setlength{\abovecaptionskip}{-5pt}
  \setlength{\belowcaptionskip}{-10pt}
  \includegraphics[width=0.5\textwidth]{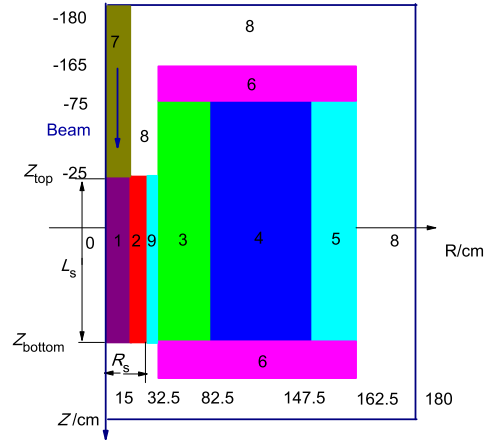}\\
  \caption{RZ view of modified ADS benchmark. The different regions are:(1)Target; (2)Buffer; (3)Core1; (4)Core2; (5)Core3; (6)Reflector; (7)Beam pipe; (8)Moderator; (9)Void.}\label{Fig:5}
\end{figure}

The size of subcritical core changes with the radius of spallation region. It was proved that proton efficiency decrease when radius of spallation region (called target in \cite{Seltborg}) increases due to the soft and leakage of neutrons \cite{Seltborg}. Changing the size of the subcritical core will bring to change of several factors that influence the proton efficiency. In this work, as is shown in \hyperref[Fig:5]{Fig. \ref*{Fig:5}}, a "void" region is inserted between the spallation region and the subcritical core to study the effect of backward neutrons. Changing the thickness of this "void" region, we could simulate different $R_{\rm{s}}$ without changing other parameters, such as the size of the subcritical core. This "void" region does not exist in practical design of ADS, but it is useful to study the effect of backward neutrons.

To study the effect of backward and radial neutrons on proton efficiency with ADS model, parameters of IAEA ADS benchmark were used, as listed in \hyperref[Tab:2]{Table \ref*{Tab:2}}; $L_{\rm{s}}=100\ \rm{cm}$. We use the target model in \hyperref[Fig:3]{Fig. \ref*{Fig:3}} to study the number of backward neutrons and the soft of radial neutrons when $R_{\rm{s}}$ changes. Four different thickness of "Void" region were chosen to get $R_{\rm{s}}=12,18,24,30\ \rm{cm}$. The result is listed in \hyperref[Tab:4]{Table \ref*{Tab:4}}. Remember that in all simulations with FLUKA, region subcritical core is filled with vacuum.
\begin{table}[hbt]
  \centering
  \setlength{\abovecaptionskip}{-5pt}
  \caption{Characteristic of spallation neutrons, $1\sigma<1.0\%$}\label{Tab:4}
  \begin{tabular}{lllll}
    \hline
    \multicolumn{5}{c}{Number of spallation neutrons in different direction }\\
    \hline
    $R_{\rm{s}}/\rm{cm}$ & 12     & 18     & 24     & 30 \\
    Radial          & 20.665 & 21.838 & 21.219 & 20.199 \\
    Forward         &  0.007 &  0.034 &  0.110 &  0.263 \\
    Backward:       &  3.729 &  6.368 &  8.702 & 10.767 \\
    \ \ Survival    &  3.530 &  5.697 &  7.270 &  8.281 \\
    \ \ Leak        &  0.199 & 0.671  &  1.432 &  2.486 \\
    \hline
    \multicolumn{5}{c}{Ratio of high-energy Radial neutrons}\\
    \hline
    $>100\ \rm{MeV} \%$   & 1.70  & 0.99 & 0.62  & 0.41 \\
    $>20\ \rm{MeV} \%$    & 1.26  & 0.86 & 0.58  & 0.39 \\
    $>1\ \rm{MeV} \%$     & 59.62 & 48.3 & 38.97 & 31.04 \\
    \hline
  \end{tabular}
\end{table}

In \hyperref[Tab:4]{Table \ref*{Tab:4}}, as $R_{\rm{s}}$, we find that: (1) the number of radial neutrons changes slightly and the spectrum of radial neutrons becomes softer (ratio of neutrons with kinetic energy higher than 1 MeV decrease from 59.62\% to 31.04\%); (2) the number of backward neutrons increases greatly (from 3.729 to 10.767); (3) the number of survival neutrons increases considerably (from 3.530 to 8.821). Explanation to these results may be: most neutrons, especially high energy neutrons are produced near the top surface of target; as $R_{\rm{s}}$ increases, it is more likely that these neutrons make it to the top surface of target and then leak out in backward direction. In conclusion, when the radius of spallation region increases, number of survival neutrons becomes considerably big, which will contribute a positive effect to proton efficiency; radial neutrons become softer, which will contribute a negative effect to proton efficiency. Comparison of these two effects are discussed below.

In this simulation, material of region "moderator" in \hyperref[Fig:3]{Fig. \ref*{Fig:3}} was vacuum, not Lead; replacing the vacuum with Lead would bring a slight decrease to the number of survival neutrons due to the scattering and absorption of neutrons in the "moderator". Another simulation had been performed; for $R_{\rm{s}} =12, 18, 24, 30\ \rm{cm}$, survival neutrons decrease to 3.409, 5.098, 6.162, 6.838 per source proton. This difference does not severely affect the fact that a rather big number of backward neutrons will enter the subcritical core when the radius of spallation region is big.

\subsection{Effect of spatial and energy distribution of spallation neutrons}
\label{sec44}
\subsubsection{Optimizing target position with spatial distribution of radial neutrons}
\label{sec441}

The spatial distribution of radial neutrons may be used to optimize the axial position of target, as discussed in \hyperref[sec42]{Sec. \ref*{sec42}} . In this work, target position is determined by the top surface of target $Z_{\rm{top}}$. Optimization of $Z_{\rm{top}}$ had been performed with the following considerations:
\begin{itemize}
  \item $L_{\rm{}s}$ will be larger than the range of proton in Lead, about 100 cm for 1.0-1.5 GeV protons. In this way, $I_{\rm{r}}(Z)$ and neutron multiplicity will not change when $Z_{\rm{top}}$ changes.
  \item The bottom of target, $Z_{\rm{bottom}}$, remains at 75 cm; $Z_{\rm{top}}$ changes from -60 cm to -15 cm and $L_{\rm{}s}$ varies from 90 cm to 135 cm.
  \item All other parameters do not change, such as $R_{\rm{s}}=32.5\ \rm{cm}$ and composition of the subcritical core. $k_{\rm{eff}}$ has a little change and will be considered in the calculation of proton efficiency.
  \item $L_{\rm{s}}$ reaches its saturation, so $<S_{\rm{n}}>/{<S_{\rm{p}}>}$ does not change when $Z_{\rm{top}}$ changes. $\varphi^{*}$ is equal to $\psi^{*}$.
\end{itemize}

\begin{figure}[hbt]
  \centering
  \setlength{\abovecaptionskip}{0pt}
  \setlength{\belowcaptionskip}{0pt}
  \includegraphics[width=0.5\textwidth]{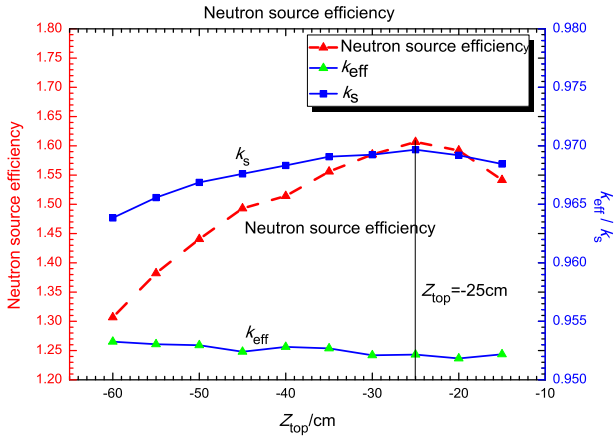}\\
  \caption{Neutron source efficiency $\varphi^{*}$, subcritical multiplication factor $k_{\rm{s}}$ and effective multiplication factor $k_{\rm{eff}}$ as a function of $Z_{\rm{top}}$ }\label{Fig:6}
\end{figure}

\begin{figure}[hbt]
  \centering
  \setlength{\abovecaptionskip}{-10pt}
  \setlength{\belowcaptionskip}{-10pt}
  \setlength{\intextsep}{-10pt plus 0pt minus 0pt}
  \includegraphics[width=0.5\textwidth]{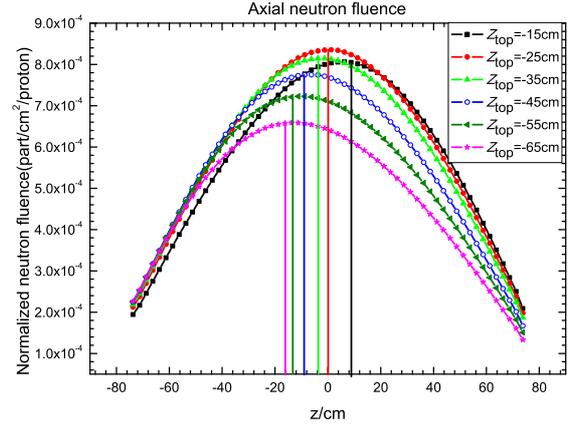}\\
  \caption{Neutron flux $\phi_{\rm{s}}(z)$ as a function of $Z_{\rm{top}}$}\label{Fig:7}
\end{figure}

The position of maximum $I_{\rm{r}}(Z)$ varies with the radius of spallation region and the energy of protons. For ADS benchmark, $R_{\rm{s}}=32.5\ \rm{cm}$, $L_{\rm{}s}=100\ \rm{cm}$, $E_{\rm{p}}=1.0\ GeV$, the maximum reaches at $Z\approx20\ \rm{cm}$. So we expected a maximum proton efficiency when $Z_{\rm(top)}=-20\ \rm{cm}$. Parameters of this simulation are listed in \hyperref[Tab:3]{Table \ref*{Tab:3}}. $\varphi^{*}$ and $\phi_{\rm{s}}(z)$ was obtained as $Z_{\rm{top}}$ changed. The result is shown in \hyperref[Fig:6]{Fig. \ref*{Fig:6}} and \hyperref[Fig:7]{Fig. \ref*{Fig:7}}.

In \hyperref[Fig:6]{Fig. \ref*{Fig:6}}, we find that both $k_{\rm{s}}$ and $\varphi^{*}$ reaches a maximum value at $Z_{\rm{top}}=-25\ \rm{cm}$. This result is a bit smaller than the position, -20 cm, we expected. This difference may be explained by the fact that the distribution of $I_{\rm{r}}(Z)$ is not perfectly symmetric, there is a "tail" at high $Z$. In \hyperref[Fig:7]{Fig. \ref*{Fig:7}}, as $Z_{\rm{top}}$ departs from $-25\ \rm{cm}$, the symmetry axis of $\phi_{\rm{s}}(z)$ departs from $z=0\ \rm{cm}$. This deviation will increase the leakage of neutrons from the subcritical core to the reflector and thus decreasing the proton efficiency.

Survival neutrons also influence $\varphi^{*}$. The number of survival neutrons will decrease when $Z_{\rm{top}}$ becomes more negative and this will bring a decrease to $\varphi^{*}$. The effect of $I_{\rm{r}}(Z)$ and survival neutrons are coupled. But when $Z_{\rm{top}}$ becomes more positive, the effect of survival neutrons becomes smaller. In conclusion, both $\phi_{\rm{s}}(z)$ and $\varphi^{*}$ support the assumption about target position. The effect of survival neutrons makes the assumption less persuasive and will be discussed below.

\subsubsection{Effect of radial and backward neutrons on proton efficiency}
\label{sec442}
Change of $R_{\rm{s}}$ will bring a change to the characteristic of radial and backward neutrons as discussed in \hyperref[sec43]{Sec. \ref*{sec43}}. These two factors are coupled. Based on simulations in \hyperref[sec43]{Sec. \ref*{sec43}}, proton efficiency $\psi^{*}$ of the modified ADS benchmark was simulated for different $R_{\rm{s}}$. Parameters of these simulations are listed in \hyperref[Tab:2]{Table \ref*{Tab:2}}, the result is listed in \hyperref[Tab:5]{Table \ref*{Tab:5}}. Changing the thickness of void region brought to a slight change of $k_{\rm{eff}}$. This change was considered in the calculation of $\psi^{*}$.
\begin{table}[htb]
  \centering
  \setlength{\abovecaptionskip}{-5pt}
  \caption{Proton efficiency varies with $R_{\rm{s}}$, $1\sigma<1.0\%$}\label{Tab:5}
  \newcommand{\rb}[1]{\raisebox{1.5ex}[0pt]{#1}}
  \begin{tabular}{lllll}
    \hline
    \multicolumn{5}{c}{Proton efficiency $R_{\rm{s}}$  }\\
    \hline
    $R_{\rm{s}}/\rm{cm}$ & 12 & 18 & 24 & 30 \\
      & 0.94828 & 0.94886 & 0.94972 & 0.95148 \\
    \rb{$k_{\rm{eff}}\pm std$} & 0.00026 & 0.00026 & 0.00025 & 0.00026 \\
    $\psi^{*}$ & 26.803 & 27.499 & 27.882 & 28.002 \\
    \hline
  \end{tabular}
\end{table}

We find that: as $R_{\rm{s}}$ increases, $\psi^{*}$ increases slightly. This result seems paradox to the conclusion that the bigger of target radius, the smaller of proton efficiency in \cite{Seltborg}. In fact,as is seen in \hyperref[Tab:4]{Table \ref*{Tab:4}}, it is the trade-off between the soft of radial neutrons and the increase of survival neutrons. Source efficiency of backward neutrons in inducing fissions is lower than these radial neutrons, but the rather big number of survival neutrons neutralizes the negative effect of radial neutrons.  This result does indicate that the effect of backward neutrons is pronounced. When radius of spallation region increase, number of neutrons produced per beam particle does increase, but these extra neutrons are not effectively used. Based on this simulation, it is difficult to draw the conclusion that the radius of spallation region should be larger; but if we need to increase radius of spallation region to satisfy other requirements in designing the system, such as decreasing the damage of high-energy spallation neutrons to structural material, we may try to make better use of these backward neutrons and avoid severely affecting the proton efficiency. At present, we may replace the moderator between the beam pipe and the subcritical core with fuels and increase the radius of spallation region; or we may design an annular target (ratio between area of  top surface and side surface is smaller than that of a cylinder target) to reduce the number of backward neutrons.

\section{Conclusion}
\label{Sec5}
 In this work, spallation region and subcritical core in ADS are combined to study effect of the spatial, energy distribution of spallation neutrons. The characteristic of radial and backward neutrons were mainly discussed.

 As for radial neutrons, neutron current surface density $I_{\rm{r}}(Z)$ was simulated with FLUKA and used to optimize the position of target. An assumption about this optimization was made and tested. As for the design of C-ADS, the considerations made in \hyperref[sec441]{Sec. \ref*{sec441}} may be used to optimize position of target. As for backward neutrons, survival neutrons were mainly discussed and found to have a considerable ratio when radius of spallation region increases. These survival neutrons have a positive effect on the proton efficiency. If we make better use of these survival neutrons, we may increase the radius of spallation region to satisfy other requirements in the design of subcritical core without affecting proton efficiency severely.
\section*{Acknowledgements}
Numerical Simulation received the support and help of Supercomputing Center of USTC. The authors acknowledge the members in the PDSP group of NSRL for their help and discussion.

\bibliographystyle{elsarticle-num}

\begin{thebibliography}{00}
\bibitem{Bauer}G.S.Bauer, \textit{Target Design and Technology for Research Spallation Neutron Sources}, Workshop on \textit{Technology and Application of Accelerator Driven Systems (ADS)},17-28 October 2005

\bibitem{Rubbia}C.Rubbia, J.A, S.Buono,\textit{Conceptual design of a fast neutron operated high power energy amplifier}, CERN/AT/95-44(ET), September 29, 1995.

\bibitem{Nifenecker}H. Nifenecker, O. M¨¦plan, S. David :\textit{Accelerator Driven Subcritical Reactors} ,Institute of Physics Publishing,London (2003)

\bibitem{Krasa}Antonin Krasa : \textit{Spallation Reaction Physics}, Chapter 1 Spallation reaction

\bibitem{Ferrari}A. Ferrari, J. Ranft, and P.R. Sala, \textit{"FLUKA: a multipar-ticle transport code"}, CERN-2005-10 (2005), INFN/TC 05/11, SLAC-R-773

\bibitem{Battistoni}G. Battistoni, S. Muraro, P.R. Sala, F. Cerutti, A. Ferrari,S. Roesler, A. Fasso`, J. Ranft,\textit{"The FLUKA code: Description and benchmarking"},Proceedings of the Hadronic Shower Simulation Workshop 2006,Fermilab 6-8 September 2006, M. Albrow, R. Raja eds.,AIP Conference Proceeding 896, 31-49, (2007)

\bibitem{RMC1} Ding SHE, Qi XU, Kan WANG et al., \textit{"RMC1.0 - Development of Monte Carlo Code for Reactor Analysis"}, Proceedings of the 18th International Conference on Nuclear Engineering (ICONE18), Xi¡¯an, China: May 17-21 (2010).

\bibitem{RMC2} Wang, K., Li, Z.G., She D., et al., \textit{Progress on RMC - A Monte Carlo Neutron Transport Code for Reactor Analysis"}, M\&C 2011, Rio de Janeiro, RJ, Brazil, May 8-12 (2011).

\bibitem{Slessarev}Slessarev I.,Tchistiakov A. IAEA ADS-BENCHMARK Results and Analysis[A]. TCM-Meeting, Madrid, 17-19 September 1997

\bibitem{Ledoux}X. Ledoux, F. Borne et al.,\textit{Spallation Neutron Production by 0.8, 1.2, and 1.6 GeV Protons on Pb Targets}, Physical review letters [0031-9007]

\bibitem{Shahbunder}Hesham Shahbunder, Cheol HoPyeon, et al. \textit{subcritical multiplication factor and source efficiency in accelerator-driven system}, Annals of Nuclear Energy Volume 37, Issue 9, September 2010, Pages 1214¨C1222

\bibitem{Seltborg}Seltborg et al., 2003 P. Seltborg, J. Wallenius, K. Tucek, W. Gudowski,\textit{Definition and application of proton source efficiency in accelerator-driven system}, Nuclear Science and Engineering, 145 (3) (2003), pp. 390¨C399

\bibitem{HEBT}Wang Xiang-qi, Luo Huan-li, et al. \textit{Distribution transformation by rotating dipole magnetic field and beam optics design for downstream of hurling magnet in C-ADS LTBT}, China Physics C(2012)

\end{thebibliography}

\end{document}